\documentstyle[aps,twocolumn,bezier]{revtex}
\begin{document}
\newcommand{\be}{\begin{equation}}
\newcommand{\ee}{\end{equation}}
\newcommand{\ba}{\begin{eqnarray}}
\newcommand{\ea}{\end{eqnarray}}
\newcommand{\by}{\begin{eqnarray*}}
\newcommand{\ey}{\end{eqnarray*}}
\newcommand{\vp}{\varphi}
\newcommand{\e}{\epsilon}
\newcommand{\ve}{\varepsilon}
\newcommand{\p}{\partial}
\newcommand{\ra}{\rightarrow}
\newcommand{\La}{\Lambda}
\newcommand{\la}{\lambda}
\newcommand{\Om}{\Omega}
\newcommand{\om}{\omega}
\newcommand{\al}{\alpha}
\newcommand{\Del}{\Delta}
\newcommand{\del}{\delta}
\newcommand{\si}{\sigma}
\newcommand{\Si}{\Sigma}
\newcommand{\f}{\frac}
\newcommand{\ti}{\tilde}
\newcommand{\pr}{\prime}
\newcommand{\ct}{\cite}
\newcommand{\ga}{\gamma}
\newcommand{\Ga}{\Gamma}
\newcommand{\scs}{\scriptstyle}
\newcommand{\sr}{\stackrel}
\newcommand{\ts}{\times}
\newcommand{\ul}{\underline}
\newcommand{\nn}{\nonumber}
\title{Oscillating Effects in the Nambu --
Jona-Lasinio Model}

\author{K.G. Klimenko}

\address{Institute for High Energy Physics, 142284 Protvino, Moscow region, Russia}

\date{\today}

\maketitle

\begin{abstract}

Phase structure of the four dimensional Nambu -- Jona-Lasinio
model has been investigated in two cases: 1) in nonsimply
connected space-time of the form $R^3\times S^1$ (space
coordinate is compactified and the length of the circle $S^1$ is
$L$) with nonzero chemical potential $\mu$ and 2) in Minkowski
space-time at nonzero values of $\mu,H$, where $H$ is the
external magnetic field. In both cases on phase portraits of the
model there are infinitly many massless chirally symmetric
phases as well as massive ones with spontaneously broken chiral
invariance. Such phase structure leads unavoidably to oscillations
of some physical parameters at $L\to\infty$ or $H\to 0$,
including magnetization, pressure and particle density of the system
as well as quark condensate and critical curve of chiral phase transitions.

\end{abstract}

\pacs{78.60.Mq,43.35.+d,95.75.Kk,25.75.Gz}

\narrowtext

\section*{1. Introduction}

This report is based on works done in collaboration with A.K.
Klimenko, M.A. Vdovichenko, A.S. Vshivtsev and V.Ch. Zhukovskii
\ct{1,2,kgk}.

The concept of dynamical chiral symmetry breaking (DCSB) plays
an essential role in elementary particle physics and quantum
field theory (QFT). In QFT this phenomenon is well observed in
Nambu -- Jona-Lasinio (NJL) type models -- four-dimensional models
with four-fermionic interactions \ct{nam}. The simplest one 
is presented by the Lagrangian, in which all spinor fields belong to
the single fundamental multiplet (one flavor case) of color $SU(N)$ group:
\be
L_\psi=\bar\psi_k i\hat\partial\psi_k
+\frac{G}{2N}
[(\bar\psi_k\psi_k)^2
+(\bar\psi_k i\gamma_5\psi_k)^2],
\ee
(here summation over color index $k=1,...,N$ is implied). Moreover, $L_\psi$ is
invariant under continuous chiral transformations
\be
\psi_k\to e^{i\theta\gamma_5}\psi_k~~;~~ (k=1,...,N).
\ee

Since there are no closed physical systems in nature, the influence of
different external factors on the DCSB mechanism is of great interest.
In these realms, special attentions have been given to analysis of the
vacuum structure of the NJL type models at nonzero temperature and chemical
potential \ct{kaw,1}, in the presence of external (chromo-)magnetic fields
\ct{klev,eb,gus}, with allowance for curvature and nontrivial space-time
topology \ct{In,2}. Combined action of external electromagnetic
and gravitational fields on DCSB effect in four-fermion field theories
were investigated in \ct{muta,od}.

In the present paper we consider the phase structure and related
oscillating effects of the four dimensional Nambu --
Jona-Lasinio model in two cases: 1) in nonsimply connected
space-time of the form $R^3\times S^1$ (space coordinate is
compactified) with nonzero chemical potential $\mu$ and 2) in
Minkowski space-time at nonzero values of $\mu,H$, where $H$ is
the external magnetic field.

\ul{\bf NJL model at $\mu\neq 0$.}
First of all let us prepare the basis for investigations in following
sections and consider phase structure of the model (1) at $\mu\neq 0$ in
Minkowski space-time.

Recall some well-known vacuum properties
of the theory (1) at $\mu=0$. The introduction of an auxiliary
Lagrangian
\be
\tilde L=\bar\psi i\hat\partial\psi-
\bar\psi (\sigma_1+i\sigma_2\gamma_5)
\psi-\frac{N}{2G}(\sigma^2_1+\sigma^2_2)
\ee
greatly facilitates the problem under consideration. (In (3) and
other formulae below we have omitted the fermionic color index $k$ for
simplicity.) On the equations of motion for auxiliary bosonic
fields $\sigma_{1,2}$ the theory (3) is equivalent to the (1) one.

From (3) one can find in the leading order of $1/N$ - expansion
the effective potential (which is the same as in 1-loop approximation)
of the model at $\mu=0$:
\ba
&\f 1N V_0&(\Sigma)=
\frac{\Sigma^2}{2G}-
\frac{1}{16\pi^2}\Biggl\{
\Lambda^4\ln
\left(1+\frac{\Sigma^2}{\Lambda^2}\right)+\Lambda^2\Sigma^2-\Biggr.\nn \\
&-&\Sigma^4\ln \left(1+\frac{\Lambda^2}{\Sigma^2}\right)\Biggl.\Biggr\},
\ea
where $\La$ is the UV cutoff parameter, $\Sigma=\sqrt{\sigma^2_1+\sigma^2_2}$.

Now one can easily see that at $G<G_c=4\pi^2/\Lambda^2$ the global
minimum point (GMP) of the potential (4) equals to zero.
Hence, in this case fermions are massless,
and chiral invariance (2) is not broken.
If $G>G_c$, then the GMP of (4) is $\Sigma_0(G,\Lambda) \not = 0$.
This mean that spontaneous breaking of
the symmetry (2) takes place. Moreover, fermions acquire mass
$M\equiv\Sigma_0 (G,\Lambda)$.

Let us now imagine that $\mu>0$. In
this case the 
effective potential $V_{\mu}(\Sigma)$ looks like \ct{1}:
\ba
V_{\mu}(\Sigma)=V_0(\Sigma)&-&
\frac{N\theta(\mu-\Sigma)}{16\pi^2}
\Biggl\{\frac{10}{3}\mu(\mu^2-\Sigma^2)^{3/2}- \Biggr.\nonumber \\
-2\mu^3\sqrt{\mu^2-\Sigma^2}&+&\Sigma^4\ln
~[~(~\mu+\sqrt{\mu^2-\Sigma^2}~)^2/\Sigma^2~]\Biggl.\Biggr\},
\ea
where $\theta(x)$ is the step function.
It follows from (5) that in the case $G<G_c$ and at arbitrary
values of chemical potential chiral symmetry (2) is not broken.
However, at $G>G_c$ the model has a rich phase structure.
On the phase portrait one can see critical curves of the second- and
first-order phase transitions, respectively.  Futhermore, there
are two tricritical points, two massive phases
with spontaneously broken chiral invariance as well
as the symmetric massless phase on the phase portrait of the
NJL model (detailed calculations of the vacuum
structure of the NJL model one can find in \ct{1}).

\section*{2. Oscillations in the $R^3\times S^1$ space-time}

It is well-known that unified theory of all forces (including
gravitation) of the nature is not constructed up to now.  Since
in early Universe the gravity was sufficiently strong, so one
should take it into account, many physicists study quantum field
theories in space-times with
nontrivial metric and topology.  At this, NJL model is the
object of special attention (see the review \ct{muta}), because the idea of
dynamical chiral symmetry breaking is the underlying concept in
elementary particle physics. There is a copious literature on this subject
\ct{In,muta,od,kim}. In particular, the investigation of four-fermion
theories in the space-time of the form $R^d\ts S^1\ts\cdots\ts S^1$ is 
of great interest \ct{kim}. The matter is that such space-time topology
occurs in superstring theories, in description of Casimir type effects
and so on.

In the present section the NJL model in the $R^3\times S^1$ 
space-time and at $\mu\neq 0$ is considered, since great amount of 
physical phemomena take place at nonzero particle density, i.e. at nonzero
chemical potential. Here space coordinate is
compactified and the circumference $S^1$ has the length $L$. For simplisity we 
study only the case with periodic boundary conditions:
$\psi(t,x+L,y,z) = \psi(t,x,y,z).$

In the leading order over $N$ the effective potential has the following form
\ct{2}:
\ba
&V&_{\mu L} (\Sigma)=V_L (\Sigma)-\f{N\la}{6\pi}
\sum^{\infty}_{n=0}\al_n
\theta (\mu - \sqrt{\Sigma^2 + (2\pi\lambda n)^2})\cdot \nn \\
&\cdot& (\mu - \sqrt{\Sigma^2 + (2\pi\lambda n)^2})^2(\mu
+2\sqrt{\Sigma^2 + (2\pi\lambda n)^2}),
\ea
where
\[
V_L (\Sigma)=V_0(\Sigma)
-\frac{2N}{\pi^2 L}\int\limits^{\infty}_{0}
dxx^2\ln[1 - e^{(-L\sqrt{x^2+\Sigma^2})}],
\]
$\al_n=2-\delta_{n0}$. Investigating potential (6) one can find \ct{2}
that at $(0.917...)G_c<G<G_c$ on the phase portrait of the model 
in the plane $(\mu,\la)$, where $\la=1/L$, there are only two massive
 nonsymmetric phases.
 In contrast, there are infinitly many massive phases in the NJL theory
if $G_c<G<(1.225...)G_c\equiv G_1$.
In both cases there are also infinitly
many symmetric massless phases in the NJL model.

In the case $G_c<G<G_1$ the boundary $\mu_c(\la)$ between symmetric
 and nonsymmetric
phases is the critical curve of second order phase transitions. This curve
oscillates at $\la\to 0$, since
\be
\mu_c(\la)\approx\f{2\pi\la_0}{\sqrt{6}}\left\{1+\f{3\la^2}{\pi^2\la^2_0}
\sum_{n=1}^\infty \f {\cos(n\pi\la_0L/\sqrt{6})}{n^2}\right\},
\ee
where $\la_0$ is defined by the relation 
$\f{\pi^2}{2G}-\f{\La^2}8 =-\f{\pi^2}6\la^2_0$.
From (7) it
follows that $\mu_{c}(\la)$ has an oscillating part, which
oscillates at $L\to\infty$ with frequency $\la_0/(2\sqrt{6})$.

Some other physical parameters of the NJL system
such as particle density, fermionic condensate and pressure
also oscillate at $L\to\infty$ \ct{2}. Let us consider in detail
oscillations of pressure and particle density in the ground state 
at $\mu>\mu_c(0)$.

This constraint means that for sufficiently large values of $L$ the
global minimum point
of the effective potential equals to zero. In this case the 
thermodynamic potential (TDP) $\Om (\mu,L)$ of the model
equals to $V_{\mu L}(0)$. Hence, using (6) we have
\ba
\Om(\mu,L)&=&V_L (0)- \nn \\
-\f{N\la}{6\pi}
\sum^{\infty}_{n=0}\al_n
\theta (\mu - 2\pi\la n)(\mu &-& 2\pi\la n)^2(\mu+4\pi\la n).
\ea
In order to extract from (8) the oscillating part one should use 
the following Poisson's summation formula \ct{lan}:
\be
\sum^{\infty}_{n=0}\alpha_n\Phi (n)
=2\sum^{\infty}_{k=0}\alpha_k
\int\limits^{\infty}_{0}\Phi (x)
\cos (2\pi kx)dx.
\ee
Then, at $L\to\infty$ the TDP $\Om (\mu,\la)$
oscillates with frequency $\mu/(2\pi)$, because it looks like \ct{2}:
\by
\Om(\mu,L)=&V&_L(0)-\f{N\mu^4}{12\pi^2}-N\sum^{\infty}_{k=0}\left
[\f{4\la^4}{\pi^2k^4} \right.\\
&-&\f{2\mu\la^3}{\pi^2k^3}\sin(\mu
kL)-\f{4\la^4}{\pi^2k^4}\cos(\mu kL)\left.\right ].
\ey
The pressure $p$ and the particle density $n$ in the ground state of the
system is defined as $p=-\p(L\Om)/\p L$ and $n=-\p\Om/\p\mu$,
respectively. Using the above expression for $\Om(\mu,L)$, we
see that these quantities also oscillates with frequency $\mu/(2\pi)$.

One can interpret the case under consideration as the ground
state of the NJL system, located between two parallel plates with
periodic boundary conditions. The force which acts on each of
plates is known as  generalized  Casimir force. Evidently, this
force is proportional to the pressure in the ground state of the
system. Hence, at nonzero chemical potential the Casimir force
of the constrained fermionic vacuum is oscillates at
$L\to\infty$.

Finally, we should like to underline that 
one can observe oscillations of some physical parameters
due to a presence in a phase structure of the NJL model of
cascades of massless as well as massive phases.

\section*{3. Magnetic oscillations at $\mu\neq 0$}

In the present section we shall study the magnetic properties of the
NJL vacuum. At $\mu =0$ this problem was considered in
\ct{klev,gus}.  It was shown in \ct{klev} that at $G>G_c$ the
chiral symmetry is spontaneously broken for arbitrary values of
external magnetic field $H$, and even for $H=0$. At $G<G_c$ the
NJL system has a symmetric vacuum at $H=0$. However, if the external
(arbitrary small) magnetic field is switched on, then for all
$G\in (0,G_c)$ one has a spontaneous breaking of initial
symmetry \ct{gus}. This is a so called effect of dynamical
chiral symmetry breaking (DCSB) catalysis by external magnetic field
(for the first time this effect was observed in the framework of
(2+1)-dimensional Gross -- Neveu model in \ct{3} and then was
explained in \ct{gus1}).

Now we turn to a more general situation when $H,\mu\neq 0$.
All the information about the vacuum structure of the NJL model is
contained in the effective potential which in 1-loop approximation and 
at $H,\mu\neq 0$ has
the following form:
\ba
V_{H\mu}(\Si)&=&V_H(\Si)-\f{NeH}{4\pi^2}\sum_{k=0}^{\infty} \al_k
\theta (\mu-s_k)\Biggl\{\mu\sqrt{\mu^2-s_k^2}\Biggr. \nn \\
&-&s_k^2\ln\left
[(\mu+\sqrt{\mu^2-s_k^2})/s_k\right ]\Biggl.\Biggr\},
\ea
where $s_k=\sqrt{\Si^2+2eHk}$, $e$ is the $N$-th part of the proton electric
charge (due to this fact, the expression (10) is only the 1-loop, but is 
not the leading order over $N$ one for effective potential),
\[
V_H(\Si)=V_0(\Si)-\f{N(eH)^2}{2\pi^2}\Bigl\{\zeta '(-1,x)
-\f 12[x^2-x]\ln x
+\f{x^2}4\Bigl.\Bigr\},
\]
$x=\Si^2/(2eH)$, $\zeta (\nu,x)$ is the generalized Riemann
zeta-function \ct{15}, $\zeta'(-1,x)$$=d\zeta(\nu,x)/d\nu|_{\nu=-1}$.

At $\mu =0$ the effective potential (10) reduces to $V_H(\Si)$. The last 
potential has nonzero global minimum point $\Si_0(H)$, such that at $G<G_c$
(only this case we shall study in the present report)
\by
\Si_0(H)\approx\f {eH}\pi \sqrt{\f G{12}}~~~~~~~~~~~~~~~\mbox&{at}&~~ H\to\infty,\\
\Si_0^2(H)\approx\f {eH}\pi\exp\{-\f 1{eH}(\f {4\pi^2}G-\La^2)\}~~\mbox&{at}&~~ H\to 0.
\ey
Therein, $\Si_0(H)$ is a monotonically increasing
function versus $H$.

Hence, at $G<G_c$ and $H=0$ the NJL vacuum is chirally symmetric one, but
arbitrary small value of external magnetic field $H$ induces the DCSB, and
fermions acquire nonzero mass $\Si_0(H)$ (the effect of magnetic
catalysis of DCSB).

In order to get a phase portrait of the model at $\mu\neq 0$ one
should find a one-to-one correspondence between points of
$(\mu,H)$-plane and global minimum points of the function (10).
It is possible to show that above critical curve
\be
\mu_c(H)=\f {2\pi}{N\sqrt{eH}} [V_H(0)-V_H(\Si_0(H))]^{1/2}
\ee
one has $(\mu,H)$-points, corresponding to the zero global
minimum of the potential (10). If $\mu<\mu_c(H)$ we have nontrivial
global minimum, which equals to $\Si_0(H)$.

Hence, at $\mu>\mu_c(H)$ ($G<G_c$) there is a region, where the vacuum
of the NJL model is symmetric one. The external magnetic
field ceases to induce the DCSB at $\mu>\mu_c(H)$
(or at sufficiently small values of magnetic field $H<H_c(\mu)$, where
$H_c(\mu)$ is the inverse function to $\mu_c(H)$).
But, under the critical curve (11) (or at $H>H_c(\mu)$) due to a
presence of external magnetic field the chiral symmetry is
spontaneously broken. Here magnetic field induces dynamical
fermion mass $\Si_0(H)$, which is not $\mu$-dependent value.
Indeed, the line (11) is the critical curve of first order phase transitions.

At first sight, properties of the 
symmetric vacuum are slightly varied, when parameters $\mu$ and $H$ are
changed. However, it is not so and in the region $\mu>\mu_c(H)$ 
we have infinitly many massless symmetric phases of the theory
as well as a variety of critical curves of the second order
phase transitions. On the experiment this cascade of phases is
identified with oscillations of such physical quantities as
magnetization and particle density.  Let us prove it.

It is well-known that a state of the thermodynamic equilibrium ($\equiv$
the ground state) of arbitrary quantum system is described by the 
thermodynamic potential (TDP) $\Om$, which is
a value of the effective potential in its global minimum point.
In the case under consideration the TDP $\Om(\mu,H)$ 
at $\mu>\mu_c(H)$ has the form
\ba
&&\Om(\mu,H)\equiv V_{H\mu}(0)=V_H(0)-\frac{NeH}{4\pi^2}\sum^{\infty}_{k=0}\alpha_k\theta
(\mu-\e_k)\cdot \nn \\
&\cdot&\{~\mu\sqrt{\mu^2-\e_k^2}-\e_k^2\ln[(\sqrt{\mu^2-\e_k^2}
+\mu)/\e_k]~\},
\ea
where $\e_k=\sqrt{2eHk}$. We shall use the following criterion
of the phase transitions: if at least one first (second) partial
derivative of $\Om(\mu,H)$ is a discontinuous function at some
point, then it is a point of the first (second) order
phase transition.

Using this criterion one can show that 
lines $l_k=\{(\mu,H):\mu=\sqrt{2eHk}\}$ ($k=1,2,...$), 
are critical lines of second order phase transitions.
Indeed, from (12) it follows that the first derivative $\p\Om/\p\mu$
is a continuous function at all lines $l_k$. However, the second
derivative $\p^2\Om/(\p\mu)^2$ has an infinite jump at each line
$l_k$:
\by
&&\f {\p^2\Om}{(\p\mu)^2}\bigg |_{(\mu,H)\to l_{k+}}-
\f {\p^2\Om}{(\p\mu)^2}\bigg |_{(\mu,H)\to l_{k-}}= \\
&&-\f {NeH\mu}{2\pi^2\sqrt{\mu^2-\e_k^2}}\bigg |_{\mu\to \e_{k+}}\ra -\infty.
\ey
So, these lines are critical curves of second
order phase transitions and we have in the NJL model an infinite set
of massless simmetric phases.

As was shown in\ct{kgk}, an infinite cascade of massless phases
is the basis for different oscillating phenomena in the NJL
model. First of all, using (9) one can present oscillating part for 
$\Om(\mu,H)$ in a manifest analytic form:
\be
\Om(\mu,H)=\Om_{mon}(\mu,H)+\Om_{osc}(\mu,H),
\ee
where ($\nu=\mu^2/(eH)$). In this formula
\[
\Omega_{mon}=V_H(0)-\f {N\mu^4}{12\pi^2}-\frac{N(eH)^2}{4\pi^3}
\int\limits^{\nu}_{0}dy \sum^{\infty}_{k=1}\frac{1}{k}P(\pi ky),
\]
\ba
&&\Om_{osc}=\frac{N\mu}{4\pi^{3/2}}
\sum^{\infty}_{k=1}
\left (\frac{eH}{\pi k}\right)^{3/2}
[Q(\pi k \nu)\cos (\pi k\nu+\pi/4)+\nn \\
&&P(\pi k \nu)\cos (\pi k\nu -\pi/4)].
\ea
Functions $P(x)$ and $Q(x)$ entering in these formulae
are connected with Fresnel's integrals $C(x)$ ¨ $S(x)$ \ct{15}:
\by
C(x)&=&\frac{1}{2}+\sqrt{\frac{x}{2\pi}}
[P(x)\sin x +Q(x)\cos x] \\
S(x)&=&\frac{1}{2}-\sqrt{\frac{x}{2\pi}}
[P(x)\cos x -Q (x)\sin x].
\ey
They have at $x\to\infty$ following asymptotics \ct{15}:
\[
P(x)=x^{-1}-\frac{3}{4}x^{-3}+...,~~~Q(x)=-\frac{1}{2}x^{-2}+\frac{15}{8}x^{-4}+...
\]
The formula (14) presents exact oscillating part of the TDP (12) for 
the NJL model at $G<G_c$. Since in the present case the TDP is proportional
to the pressure of the system, one can conclude that pressure in the NJL
model oscillates, when $H\to 0$.  It follows from (14) that
frequency of oscillations at large values of a parameter
$(eH)^{-1}$ equals to $\mu^2/2$. Then, starting from (14) one can
easily find manifest expression for oscillating parts of the particle density
$n=-\p\Om/\p\mu$ and magnetization $m=-\p\Om/\p H$. These
quantities oscillates at $H\to 0$ with the same frequency
$\mu^2/2$ and have a rather involved form, so we do not present
it here.

Magnetic oscillations in the NJL model with two light quarks
as well as dependence of dynamical quark mass on the chemical potential
at $G>G_c$ were found in \ct{16}.

The author is grateful to Prof. U. Heinz and to the Organizing Committee
of the "5th International Workshop on TFT and Their Application"
for kind invitation to attend this Workshop and hospitality as well as
 to the Deutsche Forschungsgemeinschaft
for financial support.
This work was supported in part by the Russian Fund for Fundamental 
Research, project 98-02-16690.

\vspace*{0.5cm}



\begin{thebibliography}{99}

\bibitem{1}
{\it A.S. Vshivtsev and K.G. Klimenko}, JETP Lett. {\bf 64}, 338 (1996); hep-ph/9701288;
{\it A.S. Vshivtsev, V.Ch. Zhukovskii and K.G. Klimenko}, JETP {\bf 84}, 1047 (1997).
\bibitem{2}
{\it A.S. Vshivtsev,  A.K. Klimenko and K.G. Klimenko}, Phys. Atom. Nucl. {\bf 61}, 479 (1998);
{\it M.A. Vdovichenko, A.S. Vshivtsev and K.G. Klimenko}, preprint IFVE 97-59, 
Protvino (1997) (in russian); {\it M.A. Vdovichenko, A.S. Vshivtsev and K.G. Klimenko},
JETP (1998) (to be published).
\bibitem{kgk}
{\it K.G. Klimenko}, preprint IHEP 98-56, Protvino (1998); hep-ph/9809218.
\bibitem{nam}
{\it Y. Nambu and G. Jona-Lasinio}, Phys. Rev. {\bf 122}, 345 (1961).
\bibitem{kaw}
{\it S. Kawati and H. Miyata}, Phys. Rev. {\bf D 23}, 3010 (1981);
{\it D. Ebert, Yu.L. Kalinovsky, L. M\"unchow and M.K. Volkov}, Int. J. Mod. Phys.
{\bf A 8}, 1295 (1993).
\bibitem{klev}
{\it S.P. Klevansky and R.H. Lemmer}, Phys. Rev. {\bf D 39}, 3478 (1989).
\bibitem{eb}
{\it I.A. Shovkovy and V.M. Turkowski}, Phys. Lett. {\bf B 367}, 213 (1995);
{\it D. Ebert and V.Ch. Zhukovsky}, Mod. Phys. Lett. {\bf A 12}, 2567 (1997).
\bibitem{gus}
{\it V.P. Gusynin, V.A. Miransky and I.A. Shovkovy}, Phys. Lett. {\bf B 349}, 477 (1995).
\bibitem{In}
{\it E. Elizalde, S. Leseduarte and S.D. Odintsov}, Phys. Rev. 
{\bf D 49}, 5551 (1994); 
{\it H. Forkel}, Phys. Lett. {\bf B 280}, 5 (1992); 
\bibitem{muta}
{\it T. Inagaki, T. Muta and S.D. Odintsov}, Progr. Theor. Phys. Suppl.
{\bf 127}, 93 (1997).
\bibitem{od}
{\it D.M. Gitman, S.D. Odintsov and Yu.I. Shil'nov}, Phys. Rev. {\bf D 54}, 2964 (1996);
{\it B. Geyer, L.N. Granda and S.D. Odintsov}, Mod. Phys. Lett. {\bf
A 11}, 2053 (1996); 
\bibitem{kim}
{\it D.K. Kim, Y.D. Han and I.G. Koh}, Phys. Rev. {\bf D 49}, 6943 (1994);
{\it D.K. Kim and I.G. Koh}, Phys. Rev. {\bf D 51}, 4573 (1995);
{\it K.G. Klimenko, A.S. Vshivtsev and B.V. Magnitskii}, JETP Lett. {\bf 61}, 871 (1995).
\bibitem{lan}
{\it L.D. Landau and E.M. Lifshitz}, Statistical Physics,
Nauka, Moscow (1976) (in russian).
\bibitem{3} 
{\it K.G. Klimenko}, Theor. Math. Phys. {\bf 89}, 1161 (1992);
{\it K.G. Klimenko}, Z.Phys. {\bf C 54}, 323 (1992);
{\it K.G. Klimenko, A.S. Vshivtsev and B.V. Magnitsky}, Nuovo Cim.
{\bf A 107}, 439 (1994).
\bibitem{gus1}
{\it V.P. Gusynin, V.A. Miransky and I.A. Shovkovy}, Phys. Rev.
Lett. {\bf 73}, 3499 (1994).
\bibitem{15}
{\it H. Bateman and A. Erdeyi}, Higher Transcendental Functions,
McGrawHill, New York (1953).
\bibitem{16}
{\it D. Ebert and A.S. Vshivtsev}, Van Alphen-de Haas effect for
dense cold quark matter in a homogeneous magnetic field, preprint 
HUB-EP-97/92, hep-ph/9806421.

\end{thebibliography}
\end{document}